\documentclass{isma-conf-2010}

\pdfoutput=1

\newif\ifbiblatex 
\ifbiblatex
  \usepackage[style=authoryear]{biblatex}
  \bibliography{example-refs}
  \renewbibmacro*{in:}{}% remove "In: " before the journal title
\else
  \usepackage{natbib}
  \bibpunct{(}{)}{,}{a}{}{;}
  \let\autocite\citep
\fi

\usepackage{microtype}% <= this package dramatically improves 
\usepackage[it]{subfigure}
\usepackage{epsfig}

\title{Synthetic description of the piano soundboard mechanical mobility}
\author{%
  Kerem Ege and Xavier Boutillon}
\affiliations{Laboratory for the Mechanics of Solids UMR 7649, École polytechnique, 91128 Palaiseau Cedex, France\\\emph{kerem.ege@polytechnique.edu}}

\begin{document}
\twocolumn[{%
  \maketitle
  \pacs{43.75.Mn ; 43.75.Zz ; 43.40.At}
  \begin{abstract}
An expression of the piano soundboard mechanical mobility (in the direction normal to the soundboard) depending on a small number of parameters and valid up to several kHz is given in this communication. Up to 1.1~kHz, our experimental and numerical investigations confirm previous results showing that the soundboard behaves like a homogeneous plate with isotropic properties and clamped boundary conditions. Therefore, according to the Skudrzyk mean-value theorem \autocite{SKU1980}, only the mass of the structure $M$, the modal density $n(f)$, and the mean loss factor $\eta(f)$, are needed to express the average driving point mobility. Moreover, the expression of the envelope - resonances and antiresonances - of the mobility can be derived, according to \autocite{LAN1994}. We measured the modal loss factor and the modal density of the soundboard of an upright piano in playing condition, in an anechoic environment. The measurements could be done up to 2.5 kHz, with a novel high-resolution modal analysis technique (see the ICA companion-paper, \cite{EGE2010_1}). Above 1.1 kHz, the change in the observed modal density together with numerical simulations confirm Berthaut's finding that the waves in the soundboard are confined between adjacent ribs \autocite{BER2003}. Extending the Skudrzyk and Langley approaches, we synthesize the mechanical mobility at the bridge up to 2.5 kHz. The validity of the computation for an extended  spectral domain is discussed. It is also shown that the evolution of the modal density with frequency is consistent with the rise of mobility (fall of impedance) in this frequency range and that both are due to the inter-rib effect appearing when the half-wavelength becomes equal to the rib spacing. Results match previous observations by \cite{WOG1980}, \cite{CON1996_2}, \cite{GIO1998}, \cite{NAK1983} and could be used for numerical simulations for example. This approach avoids the detailed description of the soundboard, based on a very high number of parameters. However, it can be used to predict the changes of the driving point mobility, and possibly of the sound radiation in the treble range, resulting from structural modifications.
  \end{abstract}
}]

\section{Introduction}
The bridge of a piano soundboard is the location where the energy of the strings is transferred to the soundboard. This transfer is ruled by the end condition of the strings which is described here by a 1-D mechanical mobility or admittance $Y(\omega)=V(\omega)/F(\omega)$. The purpose of this communication is to give an expression of the piano soundboard mechanical mobility depending on a small number of parameters and valid up to several kHz. In the first section we describe the coupling, and a bibliographical review of the published measurements is given. The synthetic description derived from Skudrzyk's and Langley's work is then introduced and applied to an upright piano: the average driving point mobility and its envelope are expressed with only the modal density $n(f)$, the mean loss factor $\eta(f)$ and the mass $M$ of the structure. The measurements of the modal density and the modal loss factors were done up to 2.5 kHz, with a novel high-resolution modal analysis technique presented in our ICA companion-paper \emph{Vibrational and acoustical characteristics of the piano soundboard} devoted to the vibration and some radiation characteristics of the soundboard~\autocite{EGE2010_1}. Finally, we present in the last section how it can be used to predict the changes of the driving point mobility, and possibly of the sound radiation in the treble range, resulting from structural modifications of the piano soundboard.

\section{Mechanical mobility}
\subsection{Coupling}
The research of a trade-off between loudness and sustain (duration) is a major issue for piano designers and manufacturers. The way by which the energy of vibration is transferred from the piano string to the soundboard depends in particular on the end conditions of the strings at the bridge. This is a classical problem of impedance matching between a source (the string) and a load (the soundboard).

The mechanical load presented to the string can be described by the admittance $Y(\omega)$ (also called mechanical mobility) at the connecting point (bridge) between the string and the soundboard. The admittance defines the relationship between the local velocity $V$ and the excitation force $F$. Since these quantities are both of vectorial nature, $Y$ is a $3\times 3$ matrix.  In principle, the reciprocal quantity -- the impedance $Z=Y^{-1}$ could be used as well. In most mechanical systems however (including musical string instruments) the force is imposed in one direction only and the other directions are left totally free. These cases are described by only three coefficients of the mobility matrix (see for example~\cite{BOU1999}).%\footnote{The description by the impedance would require the inversion of $Z$, needing all nine coefficients to be known. The impedance is suited to systems where \emph{velocities} are imposed, with possibly null-velocity conditions in some directions (blockage). This is seldom the case in musical instruments.}.
If, in addition, only one direction of motion is under investigation, only one mobility coefficient needs to be known. In what follows, the notation $Y$ means the ratio between the velocity and the force in the direction normal to the soundboard. It would be the $Y_{zz}$ coefficient of the full matrix and one should notice in the line of the previous discussion that $Z_{zz}\,\ne\,Y_{zz}^{-1}$.

The characteristic mobility of transverse waves in a piano string is always much higher than that of the soundboard. Literature more often deals with the characteristic impedance and with the impedance-like quantity $Y^{-1}$ at the bridge. The former ranges typically from $\approx10$~kg~s$^{-1}$ for the long and thick bass strings to $\approx5$~kg~s$^{-1}$ in the treble range \autocite{ASK2006_2}. The latter is almost $100$ times larger with an average low-frequency level near 10$^3$~kg~s$^{-1}$. According to Askenfelt \autocite{ASK2006_2}, piano makers have reached this value empirically since \emph{[it] gives the proper amount of coupling and decay rate suitable for musical purposes}. Pianos having a larger soundboard mobility at the bridge tend to sound harsh and to exhibit less than normal durations of tones according to Conklin \autocite{CON1996_2}. Conversely, if the mobility level falls significantly lower, the duration is longer than normal while, at the same time, the output seems subnormal.

\subsection{Measurements at the bridge -- A bibliographical review (Wogram, Nakamura, Conklin, Giordano)}\label{sec:mobility}
\begin{figure}
\centering
 \includegraphics[width=0.55\linewidth]{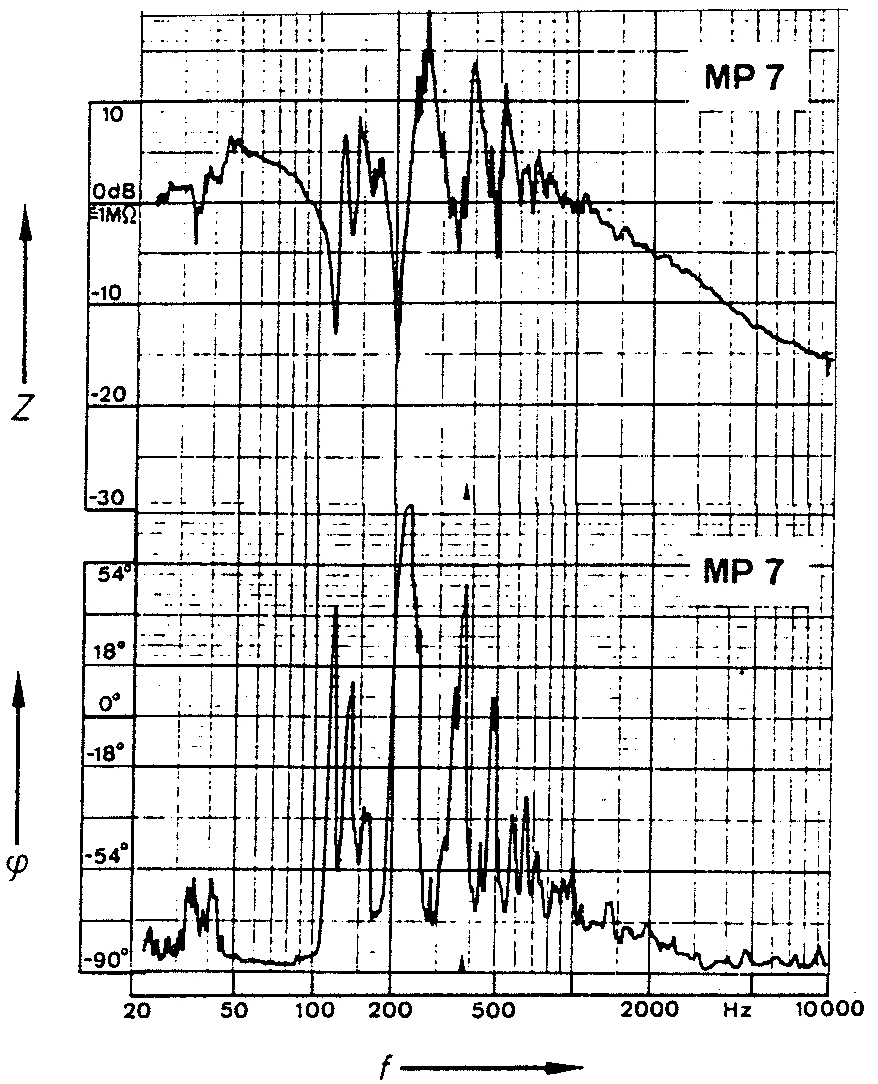}   
 \caption[Wogram]{Magnitude of the impedance $|Z|$ at bridge , and its phase $\varphi$ measured at terminating point for strings of $\mathbf{F\sharp_4}$ (key n$^\circ$~46) of an upright piano in playing situation, after \cite{WOG1980}.}
\label{fig:Wogram_point7}
 \end{figure}

\begin{figure}
\centering
\setcounter{subfigure}{0}
\subfigure[Key n$^\circ$~1 (\textbf{A$_\mathbf{0}$}\,=\,27.5~Hz)]
   {\epsfig{figure=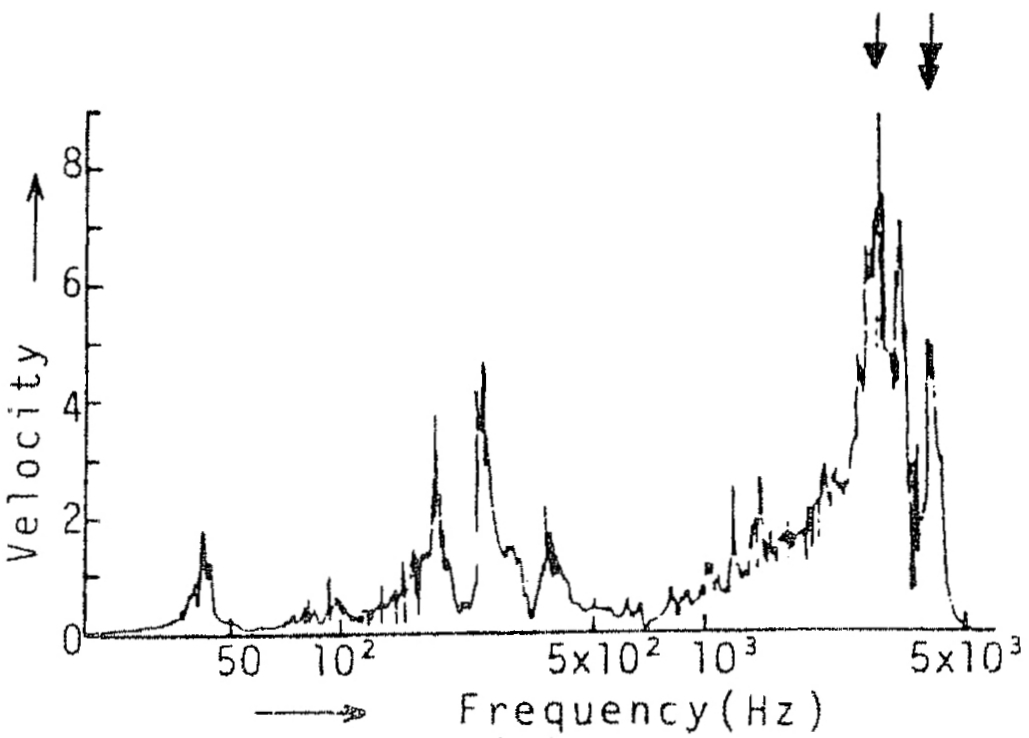,width=0.485\linewidth}}
	\hspace{0.005\linewidth}
\subfigure[Key n$^\circ$~25 (\textbf{A$_\mathbf{2}$}\,=\,110~Hz)]
  {\epsfig{figure=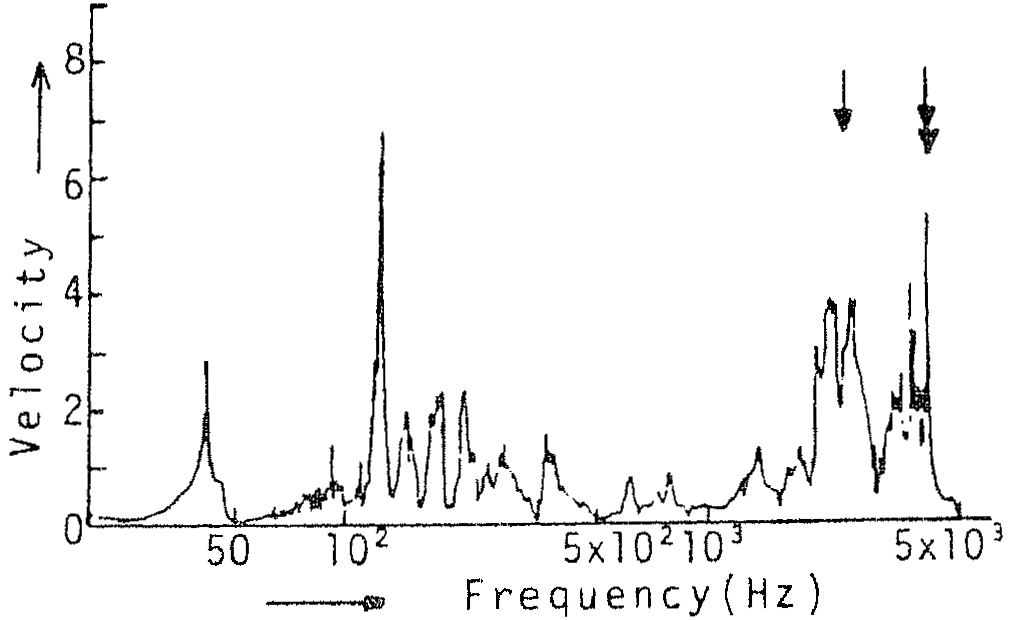,width=0.485\linewidth}}\\
\subfigure[Key n$^\circ$~27 (\textbf{B$_\mathbf{2}$}\,=\,123.5~Hz)]  
  {\epsfig{figure=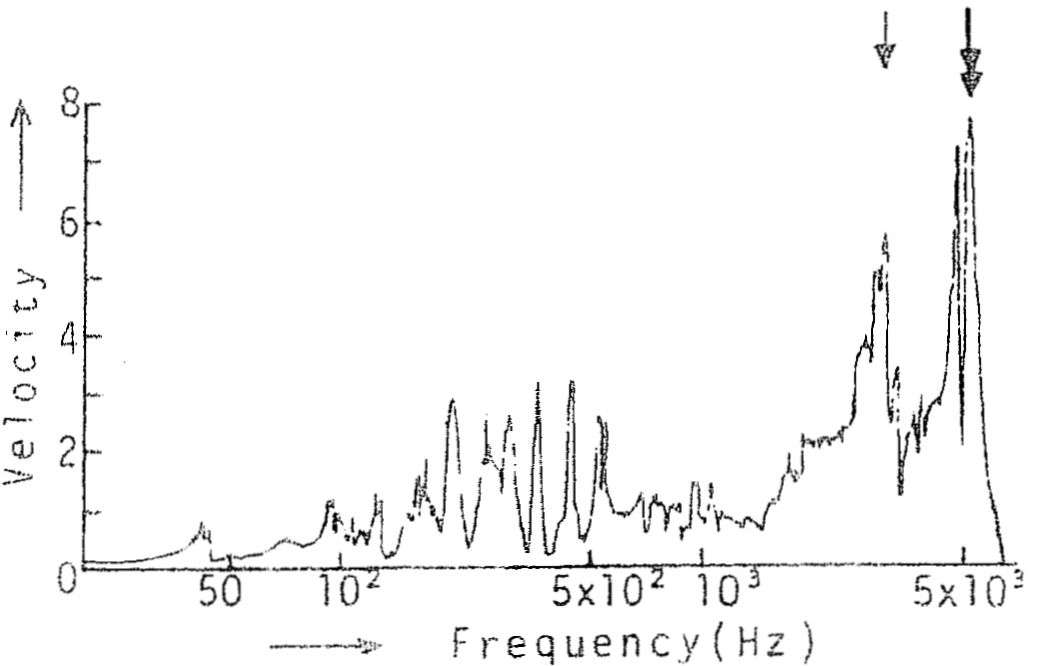,width=0.485\linewidth}}
 \hspace{0.005\linewidth}
\subfigure[Key n$^\circ$~40 (\textbf{C$_\mathbf{4}$}\,=\,261.6~Hz)]
  {\epsfig{figure=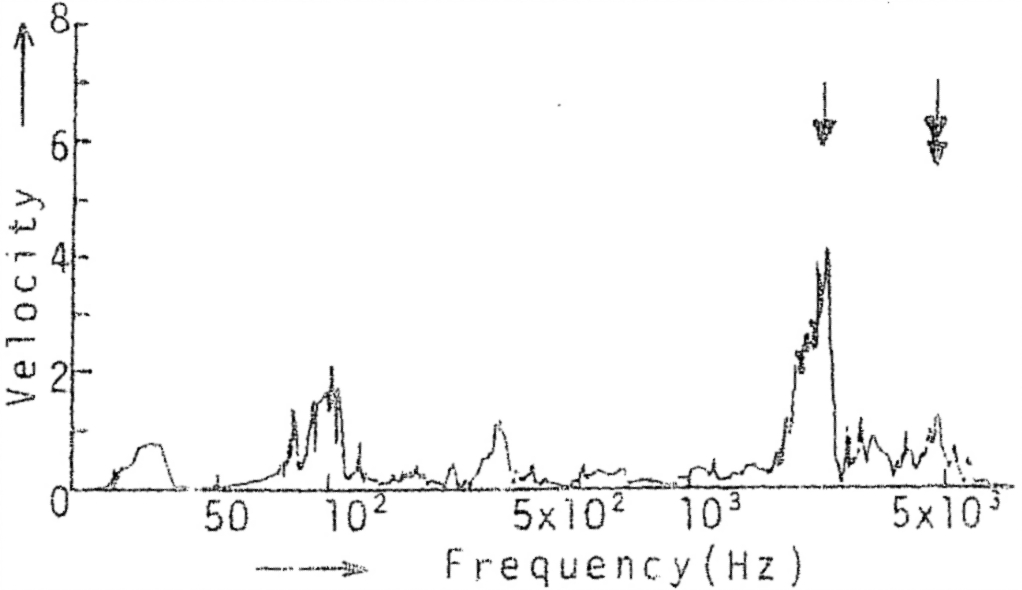,width=0.485\linewidth}}
\caption[Driving point admittance Nakamura]{Mobility at bridge at different terminating point of an upright piano in playing situation, after \cite{NAK1983}.}
\label{fig:Nakamura_admit}
\end{figure}

\begin{figure}
\centering
\includegraphics[width=0.5\linewidth]{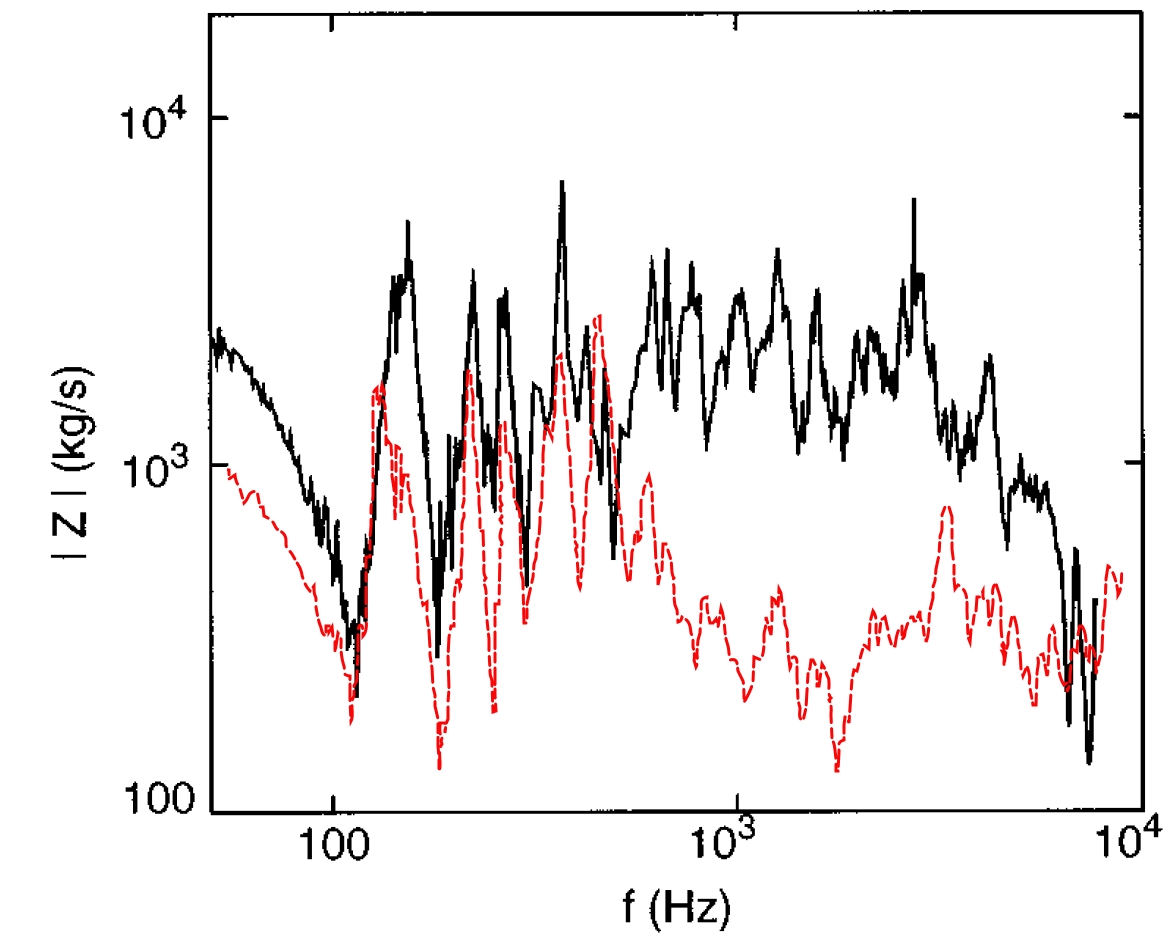}   
\caption[Giordano Compare]{Comparison of the driving point impedance measured on an upright pianos in two different locations. ---:~at bridge (terminating point for strings of $\mathbf{C_4}$ -- key n$^\circ$~40), after \cite{GIO1998}. {\color[rgb]{1,0,0} -- --} : far away from the bridge, at a mid-point between two ribs.}
\label{fig:giordano_compare}
\end{figure}

\begin{figure}[ht!]
\begin{center}
\setcounter{subfigure}{0}
  \subfigure[(0--200~Hz)]{\epsfig{figure=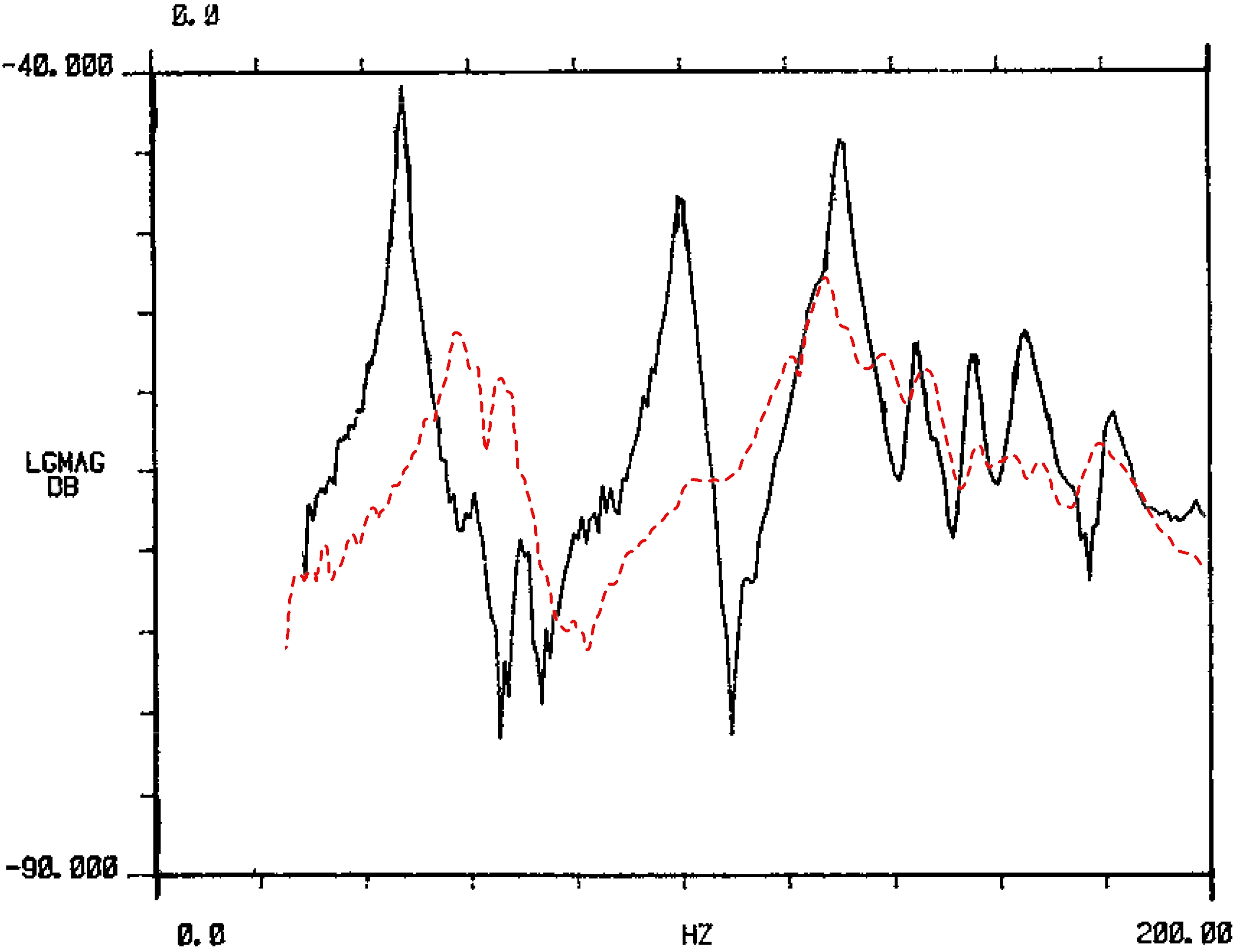,width=0.495\linewidth}}
  %\hspace{0.01\linewidth}
  \subfigure[(0--3.2~kHz)]{\epsfig{figure=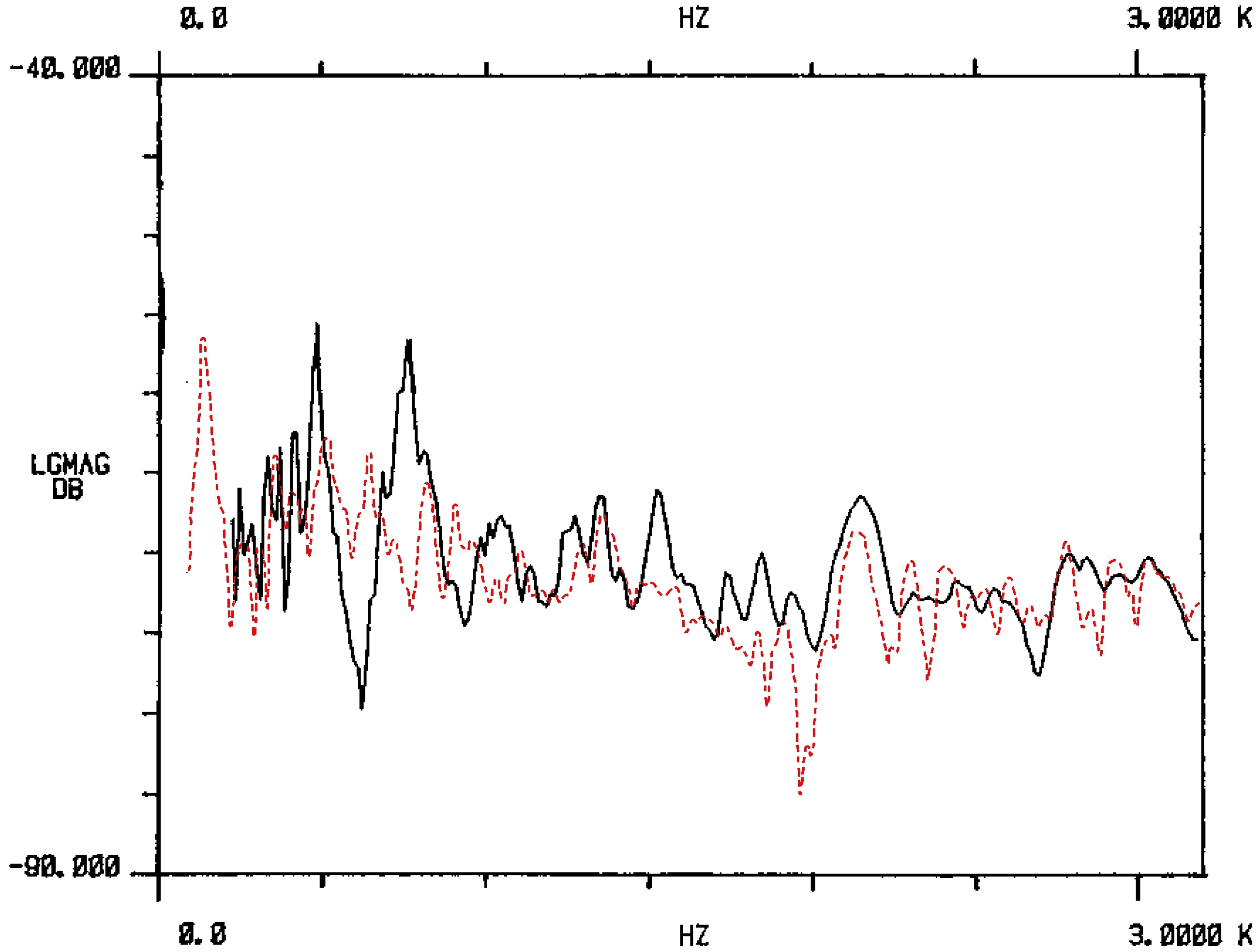,width=0.495\linewidth}}
\end{center}
\caption[Conklin mobility]{Bridge mobility (direction normal to the soundboard) of a grand piano at the terminating point of the $\mathbf{E_2}$ strings (key n$^\circ$~20), after \cite{CON1996_2}. ---:~strings and plate removed. {\color[rgb]{1,0,0} --~--}:~assembled and tuned.}
\label{fig:Conklin_mobility}
\end{figure}

Only four measurements of the admittance (or impedance) at the bridge of a piano soundboard have been published:\linebreak \cite{NAK1983} (Figure~\ref{fig:Nakamura_admit}) and \cite{CON1996_2} (Figure~\ref{fig:Conklin_mobility}) present the mobility at bridge whereas \cite{WOG1980} (Figure~\ref{fig:Wogram_point7})  and \cite{GIO1998} (Figure~\ref{fig:giordano_compare}) claim to have measured the impedance. All of these measurements are done in the direction normal to the soundboard and for upright pianos with muted strings. Conklin measured both the mobility normal to soundboard and the mobility in direction of strings of a concert grand. 

Wogram published the first impedance measurements \autocite{WOG1980}. He used an electrodynamic shaker to drive the board and an impedance head to measure at the same point the excitation force and vibration velocity. Typical results near the centre of the board are reported in figure \ref{fig:Wogram_point7}. The resonances in the soundboard motion appear as the minima of the impedance magnitude, corresponding to $\varphi=0^\circ$ in the $-\pi/2\rightarrow\pi/2$ phase transition. Between 100 and 1000~Hz, the average value of the impedance is roughly $10^3$~kg~s$^{-1}$. Above this range, $|Z|$ decreases uniformly at a rate of about 5~dB per octave to approximately $160$~kg~s$^{-1}$ at $10^4$~Hz. This rapid falloff, almost inversely proportional to frequency, appears as a measurement artefact: \emph{it has the definite appearance of some purely springy impedance which is somehow appearing in parallel with the measured one}, according to Weinreich \autocite{WEI1995}. Giordano \autocite{GIO1998} adds: \emph{it could have been caused by an effective decoupling of his impedance head from the soundboard at high frequencies}. 

Nakamura also had troubles in the high frequency range: the resonances of his driver and detector seem to have influenced heavily the coupling in this frequency range. The graphs presented for a wide frequency band (up to 5~kHz) in Figure~\ref{fig:Nakamura_admit} are the velocity normalised by the fixed driving force measured at different point of the bridges of an upright piano assembled and tuned. This quantity corresponds to the admittance at the driving point. On the graphs (Figure~\ref{fig:Nakamura_admit}), the resonances of the driver and detector are pointed out by a single arrow and a double arrow respectively. Above $1$~kHz\footnote{Note that this value is the same in Wogram's measurements.}, the mobility becomes suddenly much larger. Besides the level of resonances, this general mobility increase is, according to Nakamura, due to \emph{vibrations between ribs; in the high frequency, the ribs become the fixed edge and the inside board vibrates\footnote{Nakamura adds in the same paper that he obtained Chladni patterns where vibrations between ribs are recognised, above 1.2~kHz. Unfortunately, these figures have not been published.}}. However, Nakamura's measurements need to be reconsidered in the high frequency range.

Measurements done by Giordano \autocite{GIO1998} (Figure~\ref{fig:giordano_compare}) confirm this step-like falloff in the local impedance (or mobility increase) at high frequencies. Giordano notices that the step only occurs above approximately $2.5$~kHz when the measurement is done at bridge. 

It is interesting to notice that below the (first) impedance falloff ($\approx 700$~Hz), the average levels of the impedance measured at the bridge near a rib ($1$ to $2\cdot10^3$~kg~~s$^{-1}$) and somewhere else on the soundboard between ribs ($0.6$ to $0.7\cdot{10^3}$~kg~~s$^{-1}$) differ by a factor of 2 to 3, certainly due to the added stiffness by the bridge. The average low frequency impedance level measured at the bridge is comparable to Wogram's measurements.

\begin{figure}
\centering
\includegraphics[width=0.55\linewidth]{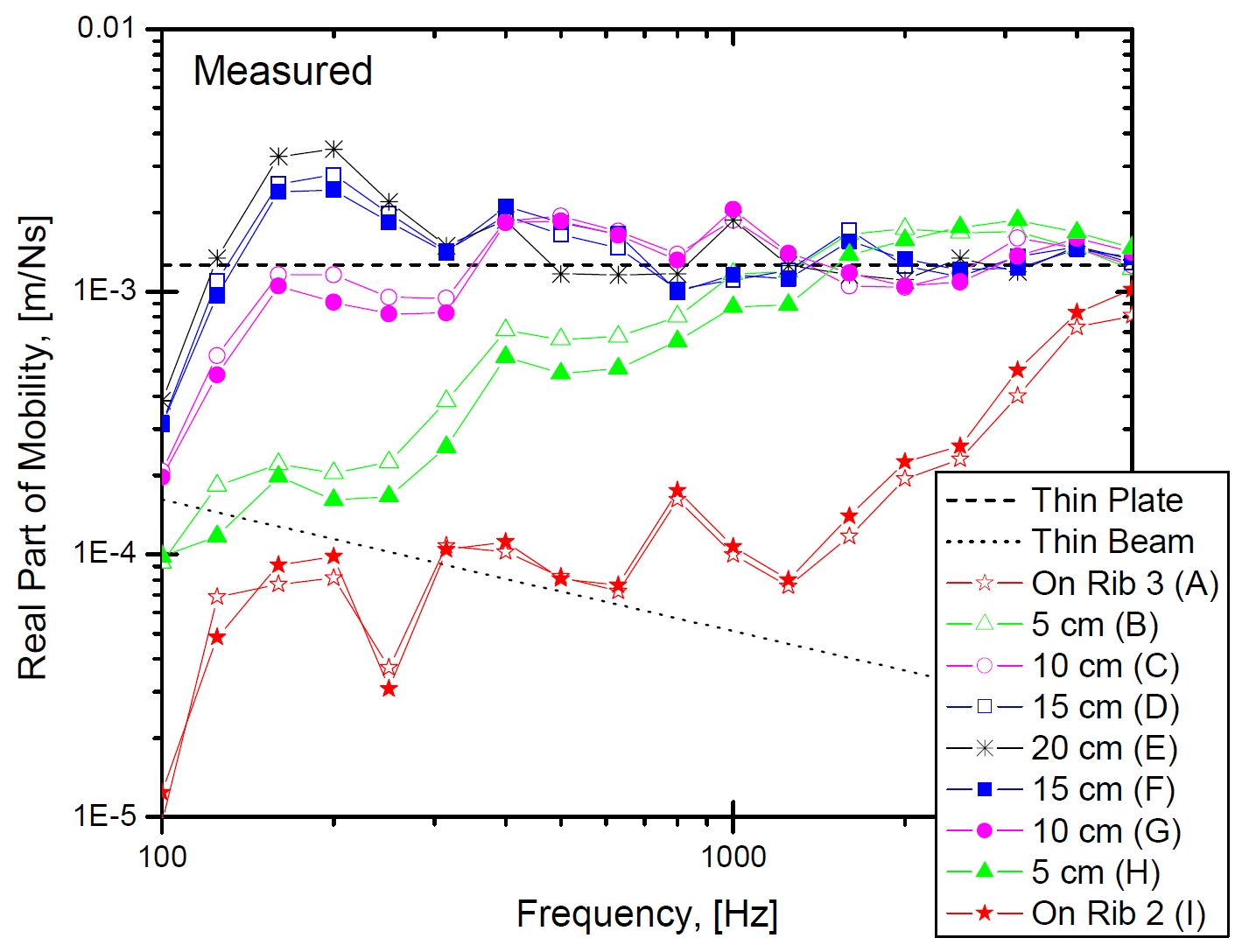}  
 \caption[nightingale]{Real part of the driving point mobility measured in different points of a ribbed plate, as a function of the distance from a rib, after \cite{NIG2006}.}
\label{fig:nightingale_mob}
\end{figure}

Nightingale \& Bosmans \autocite{NIG2006} studied the influence of the position of the driving point on the mobility of a periodic rib-stiffened isotropic plate. The space between the ribs was approximately $40$~cm. The figure~\ref{fig:nightingale_mob} points out that the real part of the mobility is a function of the distance to the nearest adjacent rib: the mobility decreases with the distance to an adjacent rib.
\begin{figure}
\centering
\includegraphics[width=0.55\linewidth]{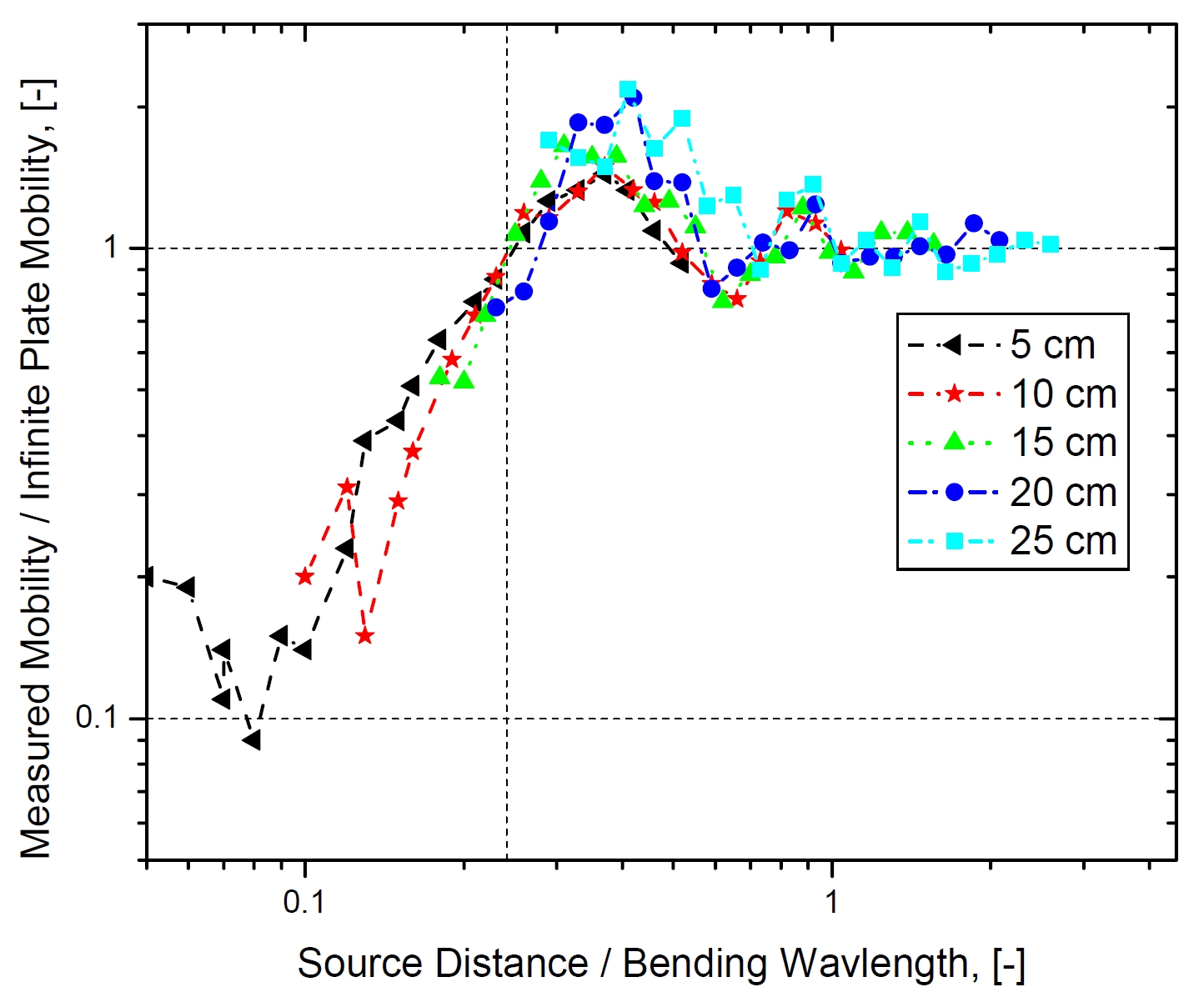}  
\caption[nightingalenorm]{Normalised mobility as a function of the normalised wave number in the guide (see text for normalising factors), after \cite{NIG2006}.}
\label{fig:nightingale_normmob}
\end{figure}

Besides, the mobility increases with frequency and tends to the mobility of an uncoupled infinite plate at high enough frequencies. The figure~\ref{fig:nightingale_normmob} presents the mobility normalised by that of an infinite plate (asymptotic value of the figure~\ref{fig:nightingale_mob}), plotted as a function of $k\,d/(2\pi)$ where $k$ is the wave number in the guide and $d$ the distance of the point of interest to the nearest rib. The ribs have almost no effect when the ratio \emph{distance to bending wavelength} is larger than 1; in other words, the ribbed plate behaves like an infinite uncoupled plate at these frequencies. When this ratio is less than $\approx0.25$ the influence of the ribs is large; the measured mobility is much less than that of the infinite plain plate.

These considerations explain why on the upright soundboard studied by Giordano, the impedance falloff between ribs appears at a much smaller frequency than when the impedance is measured on the treble bridge, close to a rib ($\approx700$~Hz for the red curve and $\approx2.5$~kHz for the black curve, in figure~\ref{fig:giordano_compare}). Moreover, in the light of the conclusions of Nigthingale \emph{et al.}, we can expect that the black and the red curves meet above 10~kHz, with a roughly constant impedance of $200$ to 300~kg~s$^{-1}$ corresponding to the characteristic impedance of the infinite plain board for bending waves.

Conklin's measurements \autocite{CON1996_2} are, to our opinion, the more accurate and reliable published measurements of a mechanical mobility at a piano bridge. Typical curves for the mobility normal to the soundboard are presented in figure~\ref{fig:Conklin_mobility}. For the sake of comparison, we superpose two sets of measurements done for the same concert grand piano. The mobility when the strings and the plate have been removed appears in solid black line and the mobility at the same point when the instrument is fully assembled and tuned in dashed red line.

Without the strings and plate, the mobility is characterised by a strong modal character up to $\approx200$~Hz. Higher in frequency, resonances are less and less pronounced and the mean value remains constant up to 3.2~kHz.

When the metal frame and strings are added, the mobility curve is substantially altered. The frequencies of the first modes is increased while the peak values are about 15~dB less. This could mean that the modification of the structure has added damping. This effect can be considered as beneficial since it reduces fluctuations in mobility, as explained in previous section. Above 1~kHz the mobility is less modified. No measurements of the mobility in the direction normal to the soundboard have been published by Conklin above $3.2$~kHz.

\begin{figure}[ht!]
\begin{center}
\setcounter{subfigure}{0}
   \subfigure[Strings and plate removed]{\epsfig{figure=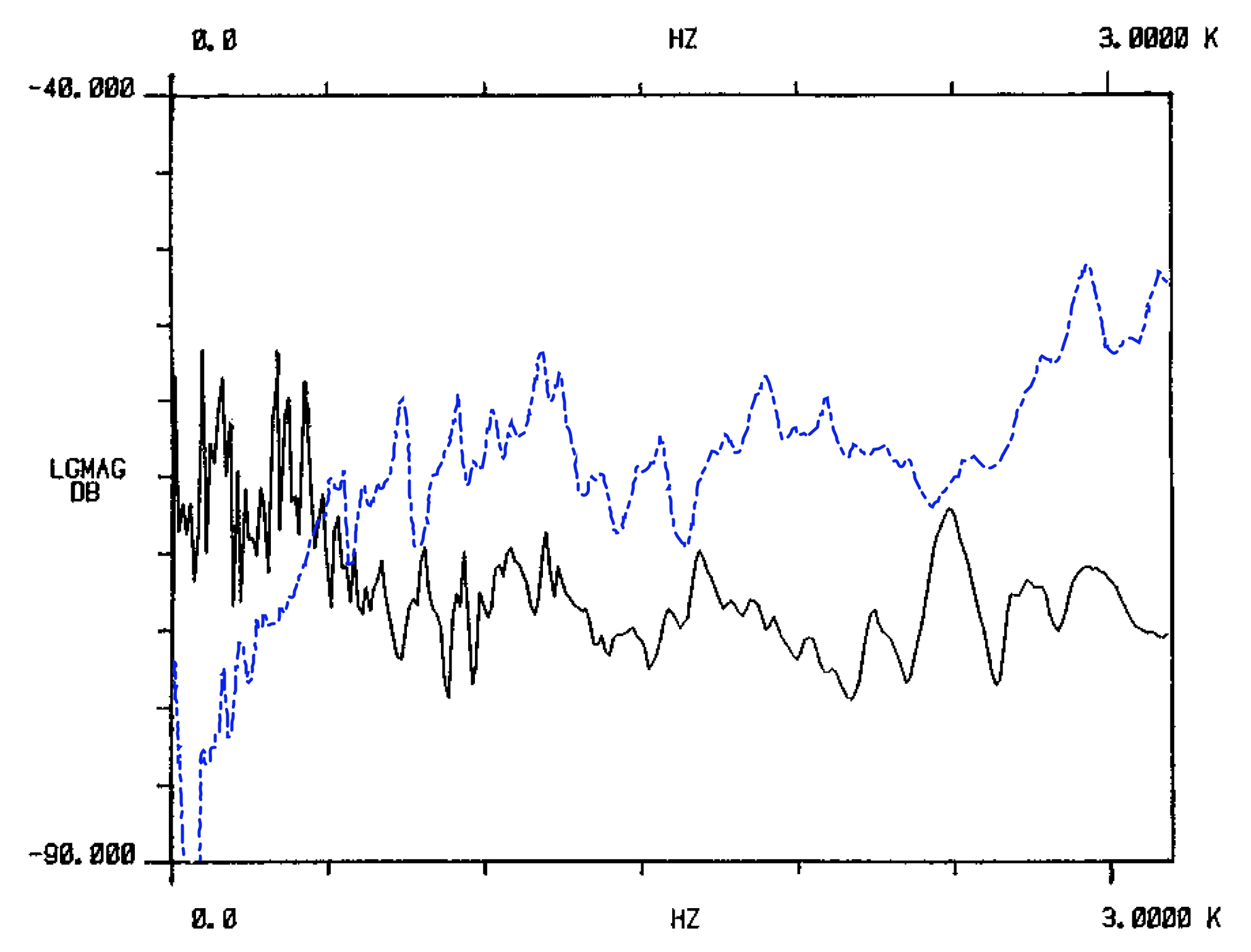,width=0.495\linewidth}}
     %\hspace{0.01\linewidth}
  \subfigure[Piano assembled and tuned]{\epsfig{figure=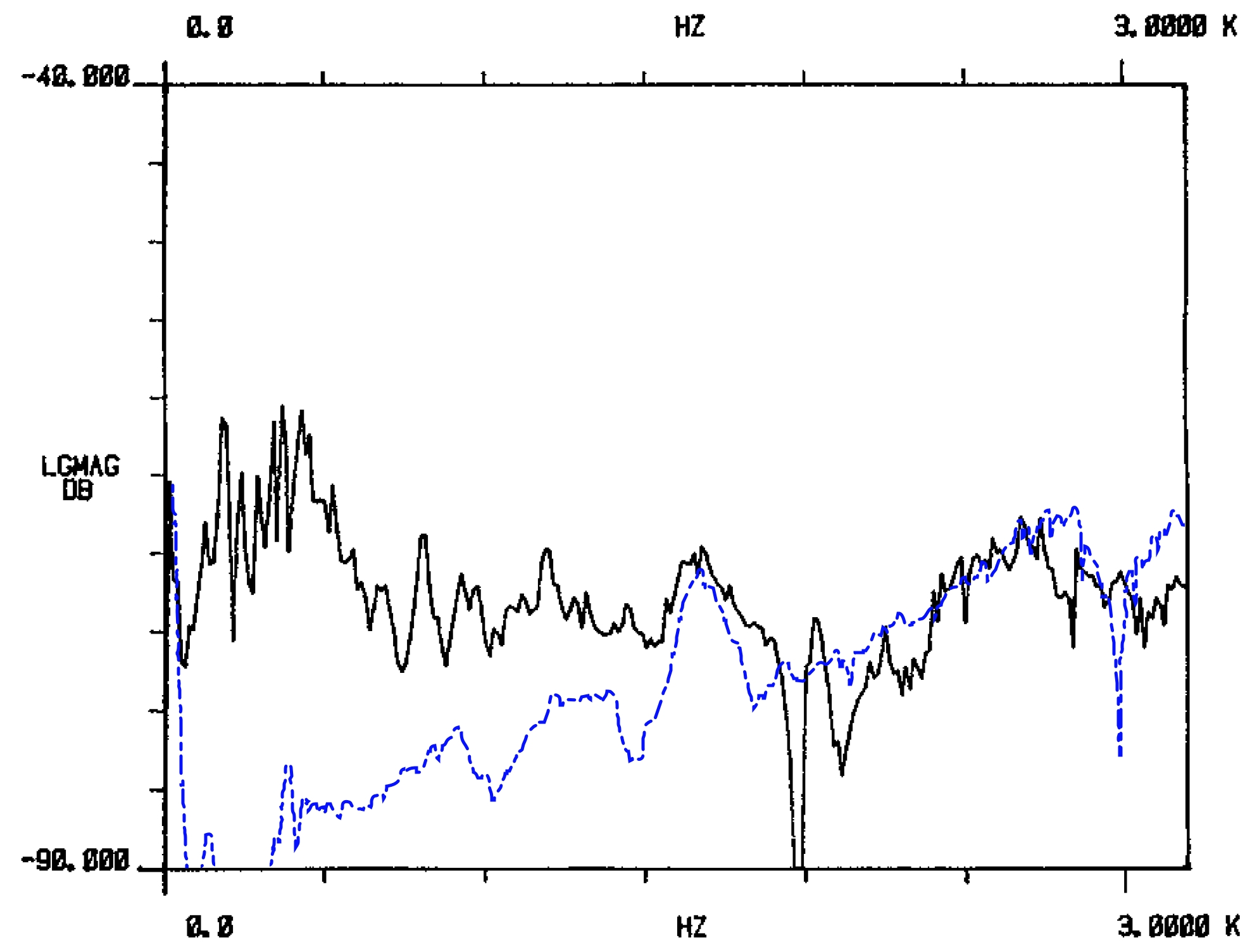,width=0.495\linewidth}}
\end{center}
\caption[Conklin longitudinalmobility]{Comparison of the transverse and longitudinal mobilities between 0 and $3.2$~kHz at the end point of the $\mathbf{C_6}$ strings (key n$^\circ$~64) of a grand piano, after \cite{CON1996_2}.  ---:~mobility in the direction normal to the soundboard. {\color[rgb]{0,0,1} --~--}:~mobility in the string direction.}
\label{fig:Conklin_longi_mobility}
\end{figure}

Conklin measured also the mobility at the bridge in the strings direction (see figure~\ref{fig:Conklin_longi_mobility}), with and without the frame and strings. Again two curves are superimposed in the figures: the mobility normal to the board (solid black line) and the "longitudinal mobility" (dashed blue line) measured at the same point.
In the treble section ($\mathbf{C_6}$ strings) and when the board is unloaded, the latter can be surprisingly larger than the former (by $10$ to $20$~dB) above $\approx1$~kHz (Fig.~\ref{fig:Conklin_longi_mobility}.a). The effect of the assembling on the longitudinal mobility is important: overall decrease of about 10--15~dB (the large longitudinal tension added by the strings to the bridge stiffens it and increases its longitudinal impedance). Thus, for a piano in playing situation, the longitudinal mobility in treble region is comparable to the mobility in the direction normal to the board for frequencies between 2-3~kHz (Fig.~\ref{fig:Conklin_longi_mobility}.b). It would be erroneous to ignore this mobility when dealing with the high-frequency tone of the piano sound. Askenfelt adds \autocite{ASK2006_2}: \emph{longitudinal string motion, which is known to influence the perception of bass notes, will thus be able to drive the soundboard rather efficiently in the high-frequency range}.

\section{Synthetic description}
The purpose here is to give an expression of the piano soundboard mechanical mobility (in the direction normal to the soundboard) depending on a small number of parameters and valid up to several kHz. 

\subsection{Analytical expression: sum of the modal contributions}
The driving point mobility $Y_\text{A}$ (at point $(x_\text{A},y_\text{A})$) of a weakly dissipative vibrating system can be expressed as the sum of the admittance of single-degree of freedom linear damped oscillators:
\begin{equation}\label{eq:drive_admitt_amort}
Y_\text{A}(\omega)=\cfrac{V_\text{A}(\omega)}{F_\text{A}(\omega)}=i\omega\,\mathop{\sum}_{\nu=1}^{+\infty}\cfrac{\Phi_\nu^2(x_\text{A},y_\text{A})}{m_\nu\,(\omega_\nu^2+i\eta_\nu\omega_\nu\omega-\omega^2)}
\end{equation}
where $m_\nu$ is the modal mass, $\eta_\nu$ is the modal loss factor, $\omega_\nu$ the modal angular frequency and  $\Phi_\nu$ the modal shape of the mode $\nu$.

\subsection{Skudrzyk mean-value theorem}
The exact expression given above is useful to study the vibratory behaviour of a structure in the low-frequency domain where only a small number of parameters is sufficient to approximate the response of the structure (usually the sum is truncated at a pulsation between 3 and 10 times the pulsation of calculus). Higher in frequency, the detailed description becomes inapplicable since the number of needed parameters is too high. Instead, we present synthetic description of the mechanical mobility at bridge based on the Skudrzyk mean-value method \autocite{SKU1980}. The method is quickly exposed here.

In the mid- and high-frequency domain, the frequency response of the structure tends to a smooth curve. The vibration can be described, ultimately, as a \emph{diffuse wavefield} (see for example~\cite{SKU1958} or \cite{LES1988}). Skudrzyk's idea, proposed in \cite{SKU1958}--\cite{SKU1968} and theorised in its final form in \cite{SKU1980} consists, in this spectral domain, in replacing the exact expression of the admittance (sum) \eqref{eq:drive_admitt_amort} by an integral. By use of the residue theorem Skudrzyk calculates the integral and shows that the real part of the admittance in high frequency may be written as a function of the ratio of the modal density $n$ and the mass of the structure $M$ only, see~\eqref{eq:Skud_admittmean}. Because $n$ and $M$ are proportional to the surface $S$, the asymptotic value of the admittance depends neither on the excitation point nor on the surface: the structure can be considered as \emph{infinite} in this frequency domain. This asymptotic value is naturally the \emph{characteristic admittance} of the structure, noted $Y_\text{C}$. By extrapolating towards the low frequencies, Skudrzyk's theory predicts the mean value and the envelope of the admittance: $G_\text{C}=\Re{(Y_\text{C})}$ is the geometric mean of the values at resonances $G_\text{res}$ and antiresonances $G_\text{ares}$. In summary, Skudrzyk's \emph{mean-value method} predicts the envelope, the mean value and the asymptotic value of the driving point admittance of a weakly dissipative vibrating structure. Contrary to statistical methods (Statistical Energy Analysis (SEA) for example, see \cite{LYO1975} e.g.), only valid in the high-frequency domain, this method gives indications on the \emph{mean} behaviour of the structure from the first resonance up to the highest frequencies. 

The principal results obtained by Skudrzyk are recalled here; for the demonstrations the reader may refer to \cite{SKU1980}. The transformation of equation \eqref{eq:drive_admitt_amort} into an integral is:
\begin{equation}\label{eq:Y_C}
Y_\text{A}(\omega)\underset{\omega\rightarrow+\infty}{\rightarrow}Y_C=\int_0^{+\infty}{\cfrac{i\omega\Phi_\nu^2(x_\text{A},y_\text{A})\,\text{d}\omega_\nu}{ m_\nu\epsilon_\nu\,(\omega_\nu^2+i\eta_\nu\omega_\nu\omega-\omega^2)}}
\end{equation}
where $\epsilon_\nu=\cfrac{\text{d}\omega_\nu}{\text{d}\nu}=2\pi\,\Delta\!f_\nu=\cfrac{2\pi}{n(f_\nu)}$ is the average modal spacing (written here for pulsations and corresponding to the inverse of the modal density $n(\omega_\nu)$). The writing of the denominator of $Y_\text{C}$ can be simplified in the hypothesis of small damping. For the oscillator $\nu$, in the weakly dissipative case ($\eta_\nu^2\ll1$), the damping term $i\eta_\nu\omega_\nu\omega$  is negligible compared to $\omega_\nu^2-\omega^2$ for all $\omega$ except on the vicinity of the resonant frequency $\omega_\nu$. The approximation introduced by \cite{SKU1958} or \linebreak \cite{CRE2005} consists in writing:
\begin{equation}
\omega_\nu^2+i\eta_\nu\omega_\nu\omega-\omega^2\approx\bar{\omega}_\nu^2-\omega^2\qquad\mbox{with\quad$\bar{\omega}_\nu^2=\omega_\nu^2(1+i\eta_\nu)$}
\end{equation} 
Given this, the equation \eqref{eq:Y_C} takes the form:
\begin{equation}\label{eq:Y_C2}
Y_C=\int_0^{+\infty}{\cfrac{i\omega\Phi_\nu^2(x_\text{A},y_\text{A})\,\text{d}\omega_\nu}{m_\nu\,\epsilon_\nu\,(\bar{\omega}_\nu^2-\omega^2)}}=G_\text{C}+i\,B_\text{C}
\end{equation}
where $G_\text{C}=\Re(Y_\text{C})$ and $B_\text{C}=\Im(Y_\text{C})$. Finally, by use of the residue theorem, the real part of the driving point admittance is given by:
\begin{equation}\label{eq:Skud_admittmean}
\Re(Y_A(\omega))\underset{+\infty}{\rightarrow}G_\text{C}=\cfrac{\pi}{2\,\epsilon_\nu\,M}=\cfrac{n(f)}{4M}
\end{equation}
In this frequency domain, the real part of the admittance depends only on the modal density and the mass of the structure. For a thin plate, the imaginary part $B_\text{C}$ vanishes at high frequency \autocite{SKU1980}:
\begin{equation}\label{eq:admitt_HF}
Y_A(\omega)\underset{+\infty}{\rightarrow} G_\text{C}=\cfrac{1}{4h^2}\sqrt{\cfrac{3(1-\nu_{xy}^2)}{E\,\rho}}
\end{equation}
written here in the isotropic case. $G_\text{C}$ is equivalent to the driving point admittance of the infinite plate~\autocite{CRE2005}. It depends neither on the frequency, nor on the surface but only on the thickness $h$ and on the elastic constants of plate: the Young's modulus $E$, the Poisson's ratio $\nu_{xy}$, and the density~$\rho$.

\subsection{Envelope}
Skudrzyk gives an approached expression of the envelope of the resonances and antiresonances. Under the assumptions of well-separated peaks and equal modal masses (peaks of the impulses responses of equal amplitudes), a single-degree of freedom damped oscillator~$\nu$ has an amplitude at resonance $f_\nu$ of:
\begin{eqnarray}\label{eq:Skud_Res}
G_{\text{res}}\approx\cfrac{1}{\eta\omega_\nu\,M}=\cfrac{n(f_\nu)}{4M}\:\cfrac{2}{\pi\mu(f_\nu)}=G_\text{C}\:\beta(f_\nu) 
\end{eqnarray}
with $\beta(f)=\cfrac{2}{\pi\mu(f)}=\cfrac{2}{\pi n(f)\eta f}$ and where the indicator $\mu(f)=n(f) \eta f$ is the modal overlap factor defined as the ratio between the half-power modal bandwidth and the average modal spacing. Generally, $\mu$ increases with frequency, and thus the amplitude of resonances decreases. In the theory of Skudrzyk, $G_\text{C}$ is the geometric mean value of the admittance (for all the frequencies) $G_\text{C}=(G_\text{res}G_\text{ares})^{1/2}$. This yields directly the amplitude of antiresonances:
\begin{equation}\label{eq:Skud_aRes}
G_{\text{ares}}\approx\cfrac{G_\text{C}}{\beta(f_\nu)}=\cfrac{n(f_\nu)}{4M}\:\cfrac{\pi\mu(f_\nu)}{2}
\end{equation}

\subsubsection{Langley calculations}
The equations above are valid for small modal overlaps. When the frequency increases, the contribution of the admittance of the \emph{neighbouring} modes needs to be considered. In that purpose, \cite{LAN1994} modifies the calculus and evaluates analytically the envelope of the sum~\eqref{eq:drive_admitt_amort}. He supposes that the resonances $f_\nu$ are regularly spaced with an average modal spacing equal to the inverse modal density at the frequency of interest, that is for the resonance $f_p$: $f_\nu-f_p=(\nu-p)/n(f_p)$. Under this assumption, the envelope of resonances $G_{\text{res}}$ becomes:
\begin{equation}
G_{\text{res}}\approx G_\text{C}\,\cfrac{\mu(f_p)}{2\pi}\,\mathop{\sum}_{j=1-p}^\infty{\cfrac{1}{j^2+(\mu(f_p)/2)^2}}
\end{equation}
%%\footnote{Langley introduit une seconde approximation au dénominateur de l'admittance. Pour $\omega=\omega_p$, et au voisinage de~$\omega_\nu$ : $~~\omega_\nu^2-\omega_p^2+i\eta_p\omega_p^2\approx(\omega_\nu+\omega_p)(\omega_\nu-\omega_p)+i\eta_p\omega_p^2\approx2\omega_p(\omega_\nu-\omega_p)+i\eta_p\omega_p^2~~$ ce qui permet d'écrire \linebreak $G_{\text{res}}\approx\Re{\left[\frac{1}{M}\,\mathop{\sum}_{\nu=1}^\infty{\frac{\eta_p\omega_p^3+2i\omega_p^2(\omega_\nu-\omega_p)}{4\omega_p^2(\omega_\nu-\omega_p)^2+\eta_p^2\omega_p^4}}\right]}$}
It is possible to calculate this sum (by extending the lower limit on the summation to $-\infty$):
\begin{equation}\label{eq:G_resLang}
G_{\text{res}}\approx G_\text{C}\:\coth{\left(\frac{\pi\mu(f_p)}{2}\right)}
\end{equation}
Similarly, supposing that the minima of admittance appear half-way between two successive resonances (that is at frequency \linebreak $f=\dfrac{f_\nu+f_{\nu+1}}{2}=f_\nu+\dfrac1{2\:n(f_\nu)}$), the envelope of antiresonances is given by:
\begin{eqnarray}
G_{\text{ares}}\approx G_\text{C}\,\cfrac{\mu(f_p)}{2\pi}\,\mathop{\sum}_{j=-\infty}^\infty{\cfrac{1}{(j-1/2)^2+(\mu(f_p)/2)^2}}\nonumber\\=G_\text{C}\:\tanh{\left(\frac{\pi\mu(f_p)}{2}\right)}\label{eq:G_aresLang}
\end{eqnarray}
For small modal overlaps, equations \eqref{eq:G_resLang} and \eqref{eq:G_aresLang} established by Langley are equivalent to the ones given by Skudrzyk: $$\coth{\left(\pi\mu/{2}\right)}\underset{0}{\sim}{2/(\pi\mu)}=\beta \quad ;\quad \tanh{\left(\pi\mu/{2}\right)}\underset{0}{\sim}\pi\mu/2={\beta}^{-1}$$ For high frequencies, these two factors have the same limit (one) and the envelope tends to $G_c$, which is consistent with the theory of Skudrzyk.

\subsubsection{Irregular natural frequency spacing}
Bidimensional structures, such as plates can present repeated resonances, degeneracy and thus irregular modal spacing. This severely degrades the accuracy of the admittance envelope given by \eqref{eq:G_resLang} and \eqref{eq:G_aresLang}. Langley introduces semi-empirical modifications in order to take into account these irregularities. The approach is based upon existing literature concerning statistical repartition of the resonances in room acoustics, \cite{BOL1946}-\cite{BOL1947} or \cite{SEP1965}. Under the assumption that the modal spacing conforms to the \emph{Poisson law}, the amplitudes of resonant frequencies of a bi-dimensional rectangular structure are given by (\cite{LAN1994}):
\begin{eqnarray}
G_{\text{res}}\approx G_\text{C}\:(1+\mu_2^{-1/2})\:\coth{\left((1+\mu_2^{-1/2})\:\frac{\pi\mu_2}{2}\right)}
\label{eq:G_resLang2}
\end{eqnarray}
where the modal overlap factor $\mu_2=[1-(L_1 L_2)^{-1}]\,\mu\,$ is modified to take into account the repeated frequencies. $\mu_2$ depends on the natural numbers $L_1$ and $L_2$ related to the aspect ratio of the rectangular structure by $L_2/L_1=L_y/L_x$. 
%\footnote{Par exemple, la pulsation propre $\omega_{mn}$ d'une plaque rectangulaire isotrope de facteur d'aspect ${L_y}/{L_x}=L_2/L_1$ sera \emph{répétée} si $L_2\,n/L_1$ et $L_1 m/L_2$ sont entiers.} (supposée rectangulaire) par $L_2/L_1=L_y/L_x$.
The amplitude of antiresonances are:
\begin{equation}
G_{\text{ares}}\approx G_\text{C}\:K\:\tanh{\left(K\frac{\pi\mu_2}{2}\right)} \label{eq:G_aresLang2}
\end{equation}
where the factor $K$ is given by the semi-empirical formula:
\begin{equation}
K=\begin{cases} 1/2.3\quad \text{si}\quad \mu_2<1\\
\cfrac{1+\mu_2-\sqrt{\mu_2}}{2.3+\mu_2-1}\quad \text{si}\quad \mu_2\geq 1
\end{cases}
\end{equation}

\section{Application on an upright piano}
\begin{figure}
\centering
\includegraphics[width=0.8\linewidth]{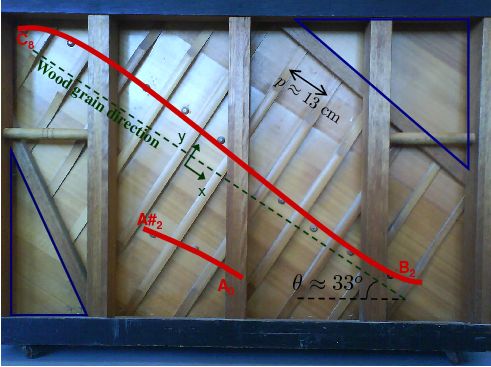}  
 \caption{Rear view of the upright piano studied.}
\label{fig:table_exp}
\end{figure}

The theory exposed in the previous section is now applied to the soundboard of an upright piano placed in a pseudo-anechoic room. The piano (see figure~\ref{fig:table_exp}) is tuned normally but strings are muted by strips of foam inserted between the strings or by woven in two or three places. A particular attention is taken not to change the mechanical mobility at bridge. The modal behaviour of the soundboard is investigated by means of a recently published high-resolution modal analysis technique \autocite{EGE2009} avoiding the frequency-resolution limitations of the Fourier transform. The method of measurements, the signal processing treatments and the extraction of the modal density and mean loss factors (up to 2.5~kHz) are exposed in detail in our ICA companion-paper \cite{EGE2010_1} devoted to the vibration and some radiation characteristics of the piano soundboard.

\begin{figure}[ht!]
\begin{center}
\includegraphics[width=1\linewidth]{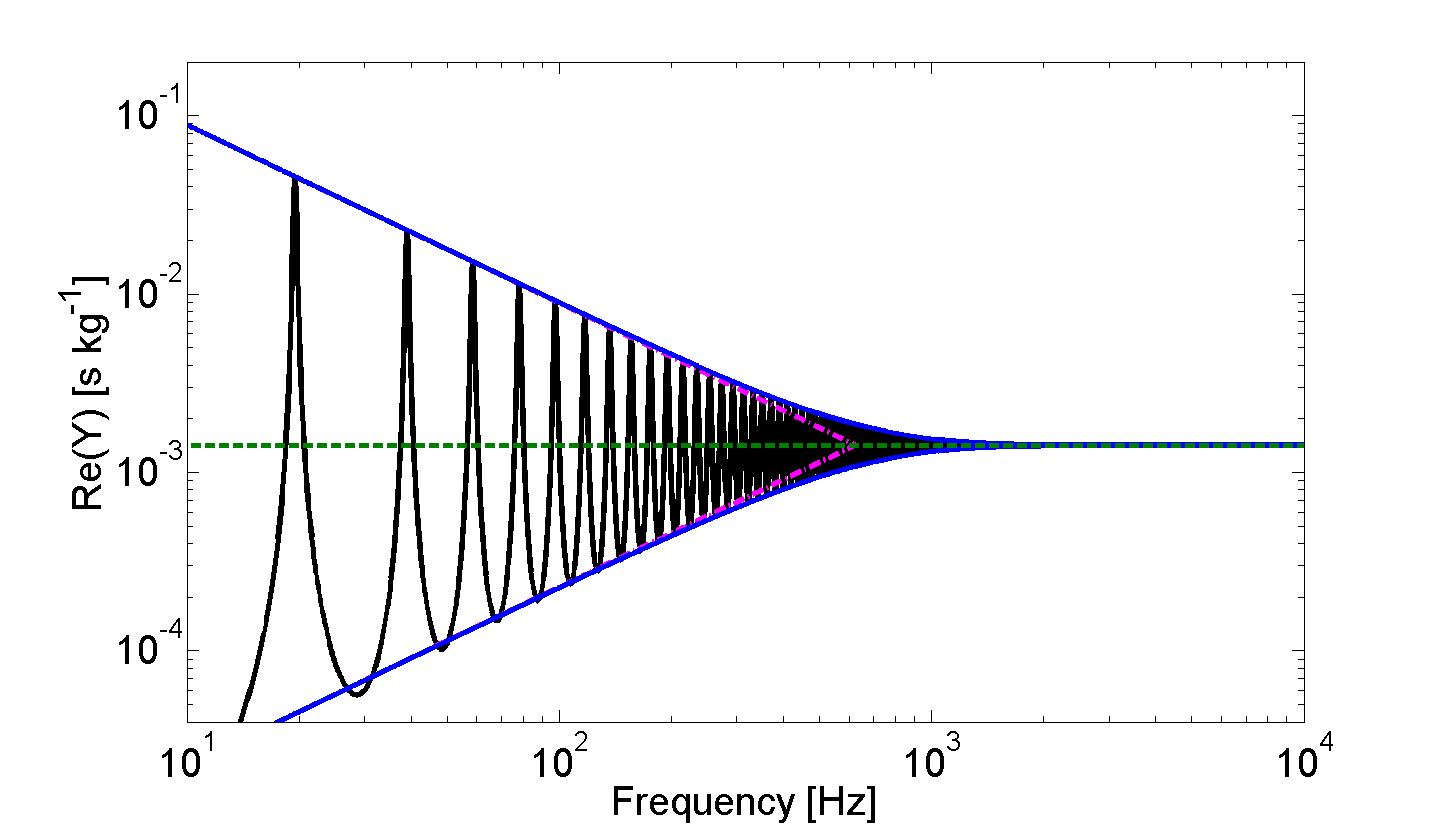}
\caption{Synthesized real part of the driving point admittance $Y_\text{A}(f)$~$(\textbf{---})$ for a weakly dissipative structure with regularly spaced resonances. The average modal spacing, the mean loss factor and the mass are measured on our upright soundboard. The characteristic admittance $G_\text{C}$~$(\textbf{{\color[rgb]{0,0.5,0} --~--}})$ is calculated with Skudrzyk's mean value method, and the envelopes of resonances and antiresonances are given by Skudrzyk~$(\textbf{{\color[rgb]{0.75,0,0.75} --$\,\cdot\,$--}})$ equations (\ref{eq:Skud_Res}-\ref{eq:Skud_aRes}) and Langley~$(\textbf{{\color[rgb]{0,0,1} ---}})$ equations (\ref{eq:G_resLang}-\ref{eq:G_aresLang}).}
\label{fig:admittoscill}
\end{center}
\end{figure}

We present in figure~\ref{fig:admittoscill} the real part of the synthesized admittance (equation \eqref{eq:drive_admitt_amort}) of the soundboard modelled as a dissipative structure where the asymptotic modal density, mean loss factor, and mass are equal to the one measured on the real structure for the frequency domain where the ribbed board behaves as a homogeneous plate (see \cite{EGE2010_1}): $n_\infty=1/19.5~\text{modes~Hz$^{-1}$}$, $\eta_\text{mean}=2\%$, $M=9$~kg. In this first calculation, the resonances are supposed regularly spaced and all the modal masses supposed equal.
The synthesized admittance tends towards the theoretical asymptote, and the envelope given by the first calculus of Langley~\eqref{eq:G_resLang} coincides for the whole spectrum with the resonances and antiresonances of the syntesized admittance. The approximation of Skudrzyk is satisfactory only for frequencies less than~$\approx200$~Hz, corresponding to a modal overlap smaller than $20\%$ (average modal spacing more than five time the half-power modal bandwidth).

\begin{figure}[ht!]
\begin{center}
\includegraphics[width=1\linewidth]{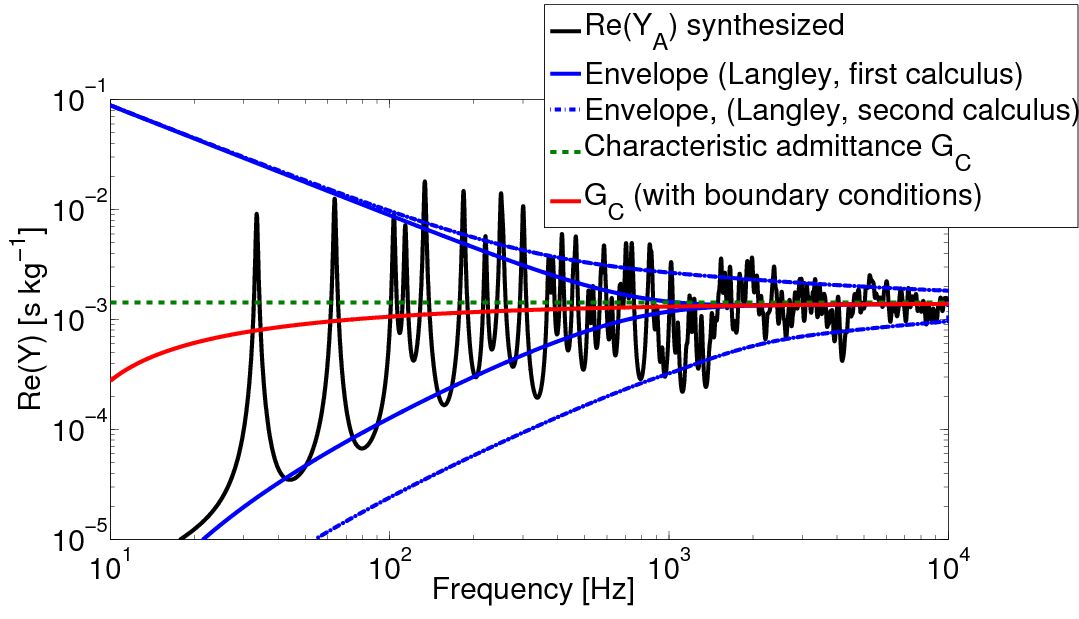}
\caption{Real part of the driving point admittance $Y_\text{A}(f)$~$(\textbf{---})$ synthesized for a point of the medium bridge of the piano soundboard studied and for simply supported boundary conditions. In order to take into account the irregularity of the modal spacing, the envelope of $\Re{(Y_\text{A})}$~$(\textbf{{\color[rgb]{0,0,1} ---}})$ is semi-empirically modified in~$(\textbf{{\color[rgb]{0,0,1} --$\,\cdot\,$--}})$ according to Langley, equations (\ref{eq:G_resLang2}-\ref{eq:G_aresLang2}). At low frequencies, the mean value $G_\text{C}$~$(\textbf{{\color[rgb]{1,0,0} ---}})$ deviates from the asymptote~$(\textbf{{\color[rgb]{0,0.5,0} --~--}})$ due to the boundary conditions.}
\label{fig:admittplaque}
\end{center}
\end{figure}

Secondly, we refined the model by considering now the structure as an isotropic rectangular plate, of constant thickness, of dimensions $L_{x}=1.39$~m, $L_{y}=0.91$~m and total mass $M=9$~kg. Indeed, up to 1.1 kHz, our experimental and numerical investigations confirm previous results showing that the soundboard behaves like a homogeneous plate with isotropic properties and clamped boundary conditions (the mechanical characteristics of the homogeneous plate are given in \cite{EGE2010_1}). For this calculus, the natural frequencies of the plate are calculated analytically (boundary conditions supposed to be simply supported). Thus, contrary to the previous case, the spectral repartition of the resonances is now irregular. The modal shapes are given by $$
\Phi_{mn}(x,y)=\sin{\left(k_{x_m}\,x\right)}\,\sin{\left(k_{y_n}\,y\right)}$$
with $m$ and $n$ natural numbers, and where the wave numbers in directions $x$ and $y$ are $k_{x_m}=\frac{m\,\pi}{L_x}$ and $k_{y_m}=\frac{n\,\pi}{L_y}$. The modal masses are equal to $M/4$. We present on figure~\ref{fig:admittplaque} the driving point admittance (equation \eqref{eq:drive_admitt_amort}) synthesized for a point of the medium bridge: here in ($4L_x/5$,$5L_y/6$).
The mean value of the mobility between 100 and 1000~Hz is approximately \linebreak $1.3\cdot~10^{-3}$~s~kg$^{-1}$  corresponding to an impedance of about 800~kg~s$^{-1}$. This value is consistent with the measurements at the bridge published by \cite{WOG1980} or \cite{GIO1998}: these authors measured a mean impedance for typical upright piano of about $10^3$~kg~s$^{-1}$ (see second section). Moreover, the fluctuations of the mobility for those frequencies are $\pm10$-$15$~dB, which is also consistent with measurements published by \cite{CON1996_2} for example. Concerning the envelope, we observe that the first calculus by Langley underestimates the amplitudes of oscillations of the mobility. The semi-empirical modifications corrects partially the envelope that becomes satisfactory around 1~kHz.	

\section{Inter-rib effect -- Structural modifications}
For frequency higher than~1100~Hz this simplest model is no more valid. Indeed, the half-wavelength at 1.1~kHz is equal to the average distance $p$ between two consecutive ribs: \emph{ribs confine the wave propagation}. The soundboard behaves as a \emph{set of waveguides} in this spectral domain. This behaviour, already found by Berthaut~\emph{et al.}~\autocite{BER2003}, is experimentally and numerically shown in our companion paper\linebreak\autocite{EGE2010_1}: for frequencies above 1.1 kHz, the modal density $n(f)$ measured on the soundboard falls significantly and the antinodes of vibration of the numerical modal shapes are localised between the ribs. A simple model of this bi-dimensional propagation media is developed from wich the modal density of the first transverse mode of the waveguide is derived. The latter takes the form (the complete calculus is presented in~\cite{EGE2010_1}):
\begin{equation}
n(\omega)=\cfrac{L_y}{\pi}\,\cfrac{\sqrt{2}\,C\omega}{\sqrt{A^2+4C\omega^2-4B}\:\left(\sqrt{A^2+4C\omega^2-4B}-A\right)^{1/2}}
\end{equation}
where $A=\cfrac{D_2+D_4}{D_3}\,k_{x_m}^2$, $B=\cfrac{D_1}{D_3}\,k_{x_m}^4$,$C=\cfrac{\rho\,h}{D_3}$ and where the $D_i$ are the constants of rigidity of spruce, considered as an orthotropic material (of main axes $x$ and $y$):\\
$D_1=E_x h^3/(12(1-\nu_{xy}\nu_{yx}))$, $D_2=\nu_{yx}E_x h^3/(6(1-\nu_{xy}\nu_{yx}))$,\\ $D_3=E_yh^3/(12(1-\nu_{xy}\nu_{yx}))$ and $D_4=G_{xy}h^3/3$.~~$E$ is the \linebreak Young's modulus, $\nu$ is the Poisson's ratio, $\rho$ the density, $h$ the plate thickness and $L_y$ the length of the waveguide. The direction $x$ is parallel to the grain of the spruce board and, thus, perpendicular to the ribs (see figure~\ref{fig:table_exp}).

We extend now the approaches of Skudrzyk and Langley for this bi-dimensional media. The mean value of the driving point mobility in this spectral domain is given by the relation~\eqref{eq:Skud_admittmean} where the modal density $n$ and mass $M$ are replaced now by the ones of the waveguide considered. Hence, for the soundboard studied, the mean value of the impedance would fall in theory from 800~kg~s$^{-1}$ before localisation of the waves to a value at 2500~Hz for example of 230~kg~s$^{-1}$ (about 3.5 times less). This value is calculated for the waveguide 2-3 (located between the second and third ribs) situated in the treble area of the instrument (upper left corner). This waveguide has a thickness $h\approx8~\text{mm}$, a length $L_y\approx55$~cm, a width (inter-rib distance) $p\approx12.8$~cm and a modal density $n\approx4\cdot 10^{-3}$~modes~Hz$^{-1}$ at 2500~Hz. 

The fall of impedance (rise of mobility) predicted by our synthetic description in the high-frequency range match previous observations by Wogram, Nakamura, Nightingale~\emph{et al.} or Giordano (see the bibliographical review given above). In particular, the value obtained theoretically for a typical waveguide is very close to the published measurements of Giordano for example (figure~\ref{fig:giordano_compare}) who measured an impedance value of about 200 to 300~kg~s$^{-1}$ in the treble area of the instrument and between two consecutive ribs. 

These results points out that the rise of mobility in this frequency range is directly linked to the inter-rib effect appearing when the half-wavelength becomes equal to the rib spacing. Thus the inter rib distance $p$ appears as a fundamental parameter in the acoustic of the instrument. If this distance $p$ is too large, the mobility at bridge will be too great in the treble range and the piano will exhibit less than normal durations of tones and a harsh sound. Conversely it $p$ is too low (too much ribs on the board) the mobility level will be small, the duration longer than normal while, at the same time, the output will seem subnormal in the treble. 

The previous paragraphs shows how the synthetic description can be used to predict the influence of a structural modification on the driving point mobility. Similarly, the modification of the thickness $h$ of the waveguide but also of the material characteristics (Young's modulus $E$, density $\rho$) may be linked to the modal density and the mass of the propagation media and thus to the mean value of the driving point mobility, thanks to the synthetic description developed.
% other structural modifications such as changes in thickness of the board, or on material characteristics (Young modulus) may be 
%
%XXX also change of material or... thickness... can change the mean value of the driving point mobility...

In order to go one step further in the analysis and envisage a possible improvement of the sound of the instrument, the construction of a soundboard on which the influence of these structural modifications may be directly measured is an absolutely necessity. We found only one experimental study, carried out by Conklin on a concert grand, where the influence of ribbing on the driving point mobility is investigated~\autocite{CON1975}. Conklin built a soundboard with $39$ ribs (more than twice the usual number), reducing the spacing $p$ to a value of $\approx$~5 to 6~cm. With this value, the first cut-off frequency (when the wavelength $\lambda=2 p$) is raised at the highest frequency of Conklin's interest, that is in his study the fundamental of the highest string of the piano: $\mathbf{C_8}\approx4186$~Hz. The height of the ribs was the same as those of a normally-designed soundboard. Their width was changed to around $1.1$~cm, approximately one half of the usual value, in order to keep almost the same stiffness and mass of the conventional board (the moment of inertia $I_\text{rib}$ of a rib that determines its stiffness is proportional to its width $a$ but varies as the cube of its height $b$: $I_\text{rib}=ab^3/12$). In his own words, Conklin's new soundboard \emph{has improved uniformity of frequency response, improved and extended high frequency response, higher efficiency at higher frequencies, and improved tone quality}. Nevertheless we believe that these conclusions need to be taken with precautions. No measure were published and the soundboard has not been commercialised. It presented surely some defects not reported by the author.

\section{Conclusion}
We have given an expression of the piano soundboard mechanical mobility (in the direction normal to the soundboard) depending on a small number of parameters and valid up to several kHz. This synthetic description is derived from Skudrzyk's and Langley's work: the mean value of the driving point mobility and its envelope are
expressed with only the modal density $n(f)$, the mean loss factor $\eta(f)$ and the mass $M$ of the structure. This theory is applied to an upright piano, from which the modal density and the modal loss factors were measured beforehand up to 2.5 kHz with a novel high-resolution modal analysis technique \autocite{EGE2010_1}. The synthetised mechanical mobility at bridge matches experimental observations and could be used for numerical simulations for example. In particular it is shown that the evolution of the modal density with frequency is consistent with the rise of mobility (fall of impedance) in this frequency range and that both are due to the inter-rib effect appearing when the half-wavelength becomes equal to the rib spacing.

This approach avoids the detailed description of the soundboard, based on a very high number of parameters. Moreover the synthetic description can be used to predict the changes of the driving point mobility, and possibly of the sound radiation in the treble range, resulting from structural modifications (changes in material, geometry, average ribs spacing, etc.).
 
 \bibliography{isma_ege_boutillon_mobility}

\end{document}